\documentclass[useAMS,usenatbib]{mn2e}

\usepackage{times}
\usepackage{graphics,epsf}
\usepackage{amsmath}                
\usepackage{amsfonts}               
\usepackage{amssymb}                
\usepackage{epsfig}                 
\usepackage{rotating}
\usepackage{color}
\usepackage{multirow}
\usepackage{natbib}

\usepackage[T1]{fontenc}
\usepackage{aecompl}

\usepackage{epstopdf}

\title[Measuring weak lensing correlations of Type Ia Supernovae]{Measuring weak lensing correlations of Type Ia Supernovae}
\author[Scovacricchi et al.]{D. Scovacricchi$^{1}$\thanks{E-mail: dario.scovacricchi@port.ac.uk}, R. C. Nichol$^{1}$, E. Macaulay$^{1,2,3}$ and D. Bacon$^{1}$\\
$^1$ Institute of Cosmology and Gravitation, University of Portsmouth, Portsmouth, UK\\
$^2$ School of Mathematics \& Physics, The University of Queensland, St. Lucia, Brisbane, 4072, Australia\\
$^{3}$ ARC Centre of Excellence for All-sky Astrophysics (CAASTRO)\\}

\begin{document}

\maketitle

\label{firstpage}

\begin{abstract}
We study the feasibility of detecting weak lensing spatial correlations between Supernova (SN) Type Ia magnitudes with present (Dark Energy Survey, DES) and future (Large Synoptic Survey Telescope, LSST) surveys. We investigate the angular auto-correlation function of SN magnitudes (once the background cosmology has been subtracted) and cross-correlation with galaxy catalogues. We examine both analytical and numerical predictions, the latter using simulated galaxy catalogues from the MICE Grand Challenge Simulation. We predict that we will be unable to detect the SN auto-correlation in DES, while it should be detectable with the LSST SN deep fields (15,000 SNe on 70 deg$^2$) at $\simeq 6\sigma$ level of confidence (assuming 0.15 magnitudes of intrinsic dispersion). The SN-galaxy cross-correlation function will deliver much higher signal-to-noise, being detectable in both surveys with an integrated signal-to-noise of $\sim100$ (up to 30 arcmin separations). We predict joint constraints on the matter density parameter ($\Omega_{\rm m}$) and the clustering amplitude ($\sigma_{8}$) by fitting the auto-correlation function of our mock LSST deep fields. When assuming a Gaussian prior for $\Omega_{\rm m}$, we can achieve a 25\% measurement of $\sigma_{8}$ from just these LSST supernovae (assuming 0.15 magnitudes of intrinsic dispersion). These constraints will improve significantly if the intrinsic dispersion of SNe Ia can be reduced.
\end{abstract}


\section{Introduction}

Type Ia Supernovae (SNe Ia) are key cosmological probes, providing some of the first evidence for an acceleration in the expansion history of the Universe \citep{riess98,perl99}. In recent years, several authors have studied the possibility of using the weak gravitational lensing of distant SNe Ia as an additional cosmological probe, providing constraints on  the growth rate of cosmic structures, which in turn can be used to constrain the contents of the Universe and alternative theories of gravity. 

Supernova (SN) lensing involves the study of the distribution of the observed SN Ia magnitude residuals (once the contribution of the background cosmology has been subtracted) as a function of redshift; this is typically referred to as the ``Hubble residuals''. Such an analysis can provide information about the gravitational perturbations along the line-of-sight, since gravitational lensing will introduce an additional non-Gaussian scatter into the SNe Ia Hubble diagram (HD). This ``one-point'' analysis of the SNe Ia magnitude distribution focuses on the possible change in the moments of the distribution (variance, skewness and kurtosis) with redshift, as predicted by the weak lensing gravitational magnification effect. 

This one-point statistical analysis was outlined in \citet{linder1988} and \citet{dodelson2005}. In recent years, \citet{marra2013} and \citet{quartin2014} have developed the technique further using their ``MeMo'' likelihood methodology to estimate the non-Gaussian behavior of the SN magnitude residuals due to weak lensing magnification. In \citet{castro2014}, they applied this technique to the JLA SN sample of \citet{betoule2014}, obtaining new constraints on $\sigma_8$ (the amplitude of density fluctuations in the Universe). \citet{Amendola2015} extended this technique to alternative cosmological scenarios, by including $\gamma$ (a measure of the growth of structure in the Universe) as a free parameter. Following different methodologies, \citet{castro2015} and \citet{macaulay2016} have investigated the constraining power of combining this one-point analysis with the magnitude effects induced by peculiar velocities. In \citet{castro2015}, they accounted for correlations in the peculiar velocities with a full covariance matrix analysis, while \citet{macaulay2016} included the effect of velocities on the magnitude moments, finding $\sigma_8 =1.072 ^{+0.50}_{-0.76}$ and $\Omega_{\rm m} = 0.278 ^{+0.011}_{-0.011}$ (68\% confidence) from the JLA data alone.  \citet{castro2016} showed that SN lensing can provide constraints on the Halo Mass Function (HMF), when combined with halos catalogues.
  
The one-point approach can be further enhanced by correlating the observed SNe Ia Hubble residuals with the matter along the line-of-sight, using galaxies as tracers of the density field \citep{kronborg2010,jonsson2010}. Such an analysis was recently performed by \cite{smith2014} for a sample of 608 SNe Ia from the SDSS-II SN Survey \citep{sako2014} and 70,631 foreground galaxies taken from the SDSS database. They found a mild correlation consistent with that expected from weak lensing (at a significance of $1.7 \sigma$). 

In this paper, we study an approach to extending these one-point statistics by considering the coherent SN brightness correlations induced by the large-scale structure in the Universe. Such structures in the foreground of distant supernovae will result in similar magnification effects being introduced into the magnitudes of neighbouring supernovae, leading to a two-point correlation, where there is an excess probability of finding magnified (or demagnified) pairs of SNe Ia, as a function of angular separation (see \citealt{hui2006} for a detailed discussion on the possible origins of magnitude spatial correlations).

A measure of the magnitude-magnitude angular correlation function would provide a direct measurement of the lensing power spectrum \citep{cooray2006_signal}, which contains information on the background cosmological expansion and the growth of structure. This will provide an independent probe for cosmology, to be combined with other techniques (e.g. galaxy shear), as well as a possible check for systematics for the upcoming wide area SN surveys.
The ideal survey to detect this correlation is deep (since lensing effects increase as we go further in redshift) and narrow, in order to provide a large number of SNe at small angular separations. However, cosmological parameter estimation using this type of surveys will suffer a non-negligible covariance, as shown by \citet{cooray2006_noise}.
We also investigate the cross-correlation between SN Hubble residuals and galaxies via analytical methods. Such correlation should be detectable with higher signal-to-noise, given the greater number density of galaxies on the sky. 

We study here the prospects for detecting the two-point angular correlation function of SN magnitudes using the Dark Energy Survey \citep[DES;][]{bernstein2012} and the Large Synoptic Sky Telescope \citep[LSST;][]{lsst_science_book}. The DES SN Survey is already underway \citep[see][]{bernstein2012,kessler2015}, while LSST will obtain first light early next decade\footnote{www.lsst.org}. In Section 2, we present the theoretical background for this work, including predictions for the signal-to-noise of the expected two-point correlation function. We investigate the likelihood of observing such correlations for both the DES and LSST surveys. In Section 3, we compare our methodology to numerical simulations to validate our approach, further investigating the possible constraints of LSST SN lensing on the constraints of the cosmological parameters $\Omega_{\rm m}$ and $\sigma_8$. We conclude the paper in Section 4, also providing for completeness a measurement of the two-point correlation function using the JLA SN sample. 

Throughout this paper, we assume a fiducial flat $\Lambda$CDM cosmology, with the matter and vacuum density of $\Omega_{\rm m}=1-\Omega_{\Lambda}=0.3$ (including the contribute of baryons $\Omega_{\rm b}=0.044)$. Where appropriate, we assume $H_0=68$ km s$^{-1}$ Mpc$^{-1}$, $\sigma_8=0.79$ and the spectral index $n_{\rm s}=0.96$, as given by the Planck cosmological analysis \citep{planck_collaboration}.

\begin{table}
\caption{Survey parameters for DES and LSST ($N_{\rm s}$ and $A_{\rm s}$ are the number of SNe and the survey area respectively).}
\begin{center}
\begin{tabular}{ l | l | l | c | c }\hline \hline
Survey & & $N_{\rm s}$  & $A_{\rm s}$ [deg$^2$]\\ \hline
DES & shallow & 1850& 24\\
 & deep & 1650& 6\\
 & hybrid &3500 & 30\\ \hline
LSST & shallow &  100k/500k/1M&20000 \\
  & deep & 10k/15k/20k& 70\\
\hline
\label{table_surveys}
\end{tabular}
\end{center}
\end{table}

\begin{table}
\caption{Galaxy surface density ($\Sigma_{\rm gal}$) and parameters ($z_0$, $\alpha$ and $\beta$) for DES and LSST galaxy survey fitting formulae (Eq. \ref{p_gal}). }

\begin{center}

{}
\begin{tabular}{ l | l | l | c | c }\hline \hline
Survey & $\Sigma_{\rm gal}\left[\frac{\rm gal}{\rm arcmin^2} \right]$  & $z_0$ &$\alpha$ & $\beta$ \\ \hline
DES & 5 & 0.7 & 2 & 1.5\\
LSST & 55 & 0.3 & 2 & 1 \\
\hline
\label{table_galsurveys}
\end{tabular}
\end{center}

\end{table}

\begin{figure*}
\includegraphics[width=.45\textwidth]{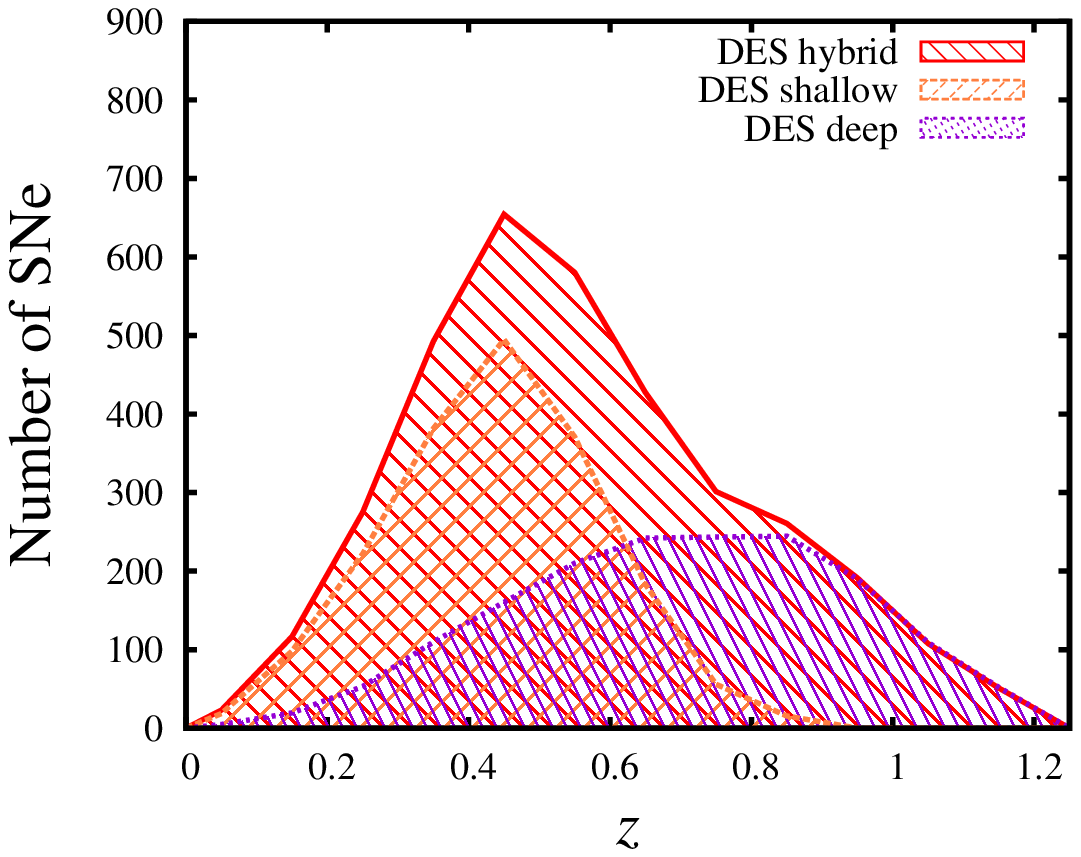}
\includegraphics[width=.45\textwidth]{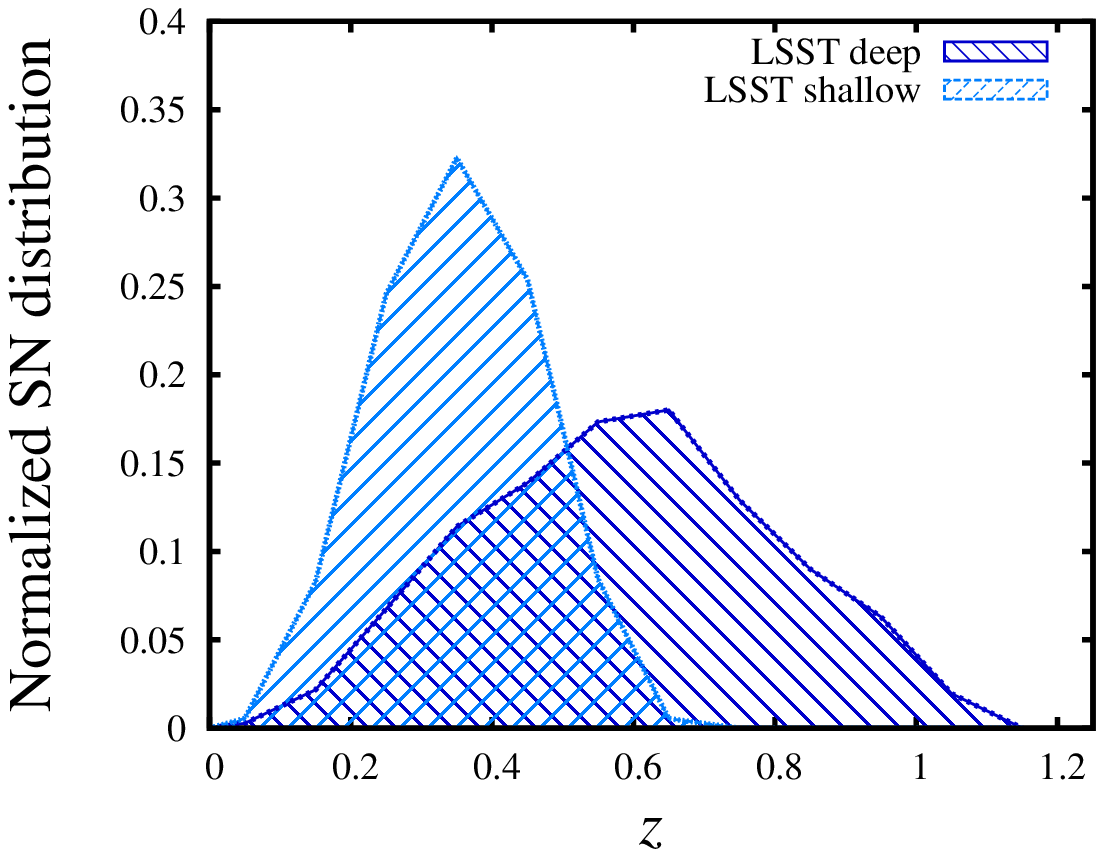}
\caption{Redshift histogram for well-observed SNe Ia from DES (left panel, $\Delta z=0.1$) and LSST (right panel). For the latter, the histograms are normalized to unit area.}
\label{DESandLSST_histograms}
\end{figure*}

\section{Analytical approach}
\label{analitycal_predictions}

\subsection{Auto-correlation function}

We present here predictions for the expected signal-to-noise of future measurement of the SN magnitude--magnitude angular correlation function. Throughout this section, we assume the line element of the first post-Newtonian order of the Minkowski metric,
\begin{equation}
ds^2 = \left(1+\frac{2 \Phi}{c^2}  \right)c^2 dt^2 -\left(1-\frac{2 \Phi}{c^2}  \right)dx^2
\end{equation}
in an otherwise Friedmann-Robertson-Walker universe. The density perturbations are well localised, their related Newtonian potential $\Phi$ is small (i.e. $\Phi \ll c^2$) and typical velocities are much smaller than the speed of light. Finally, we use the Born approximation to define the convergence $\kappa$ as an integral along the line-of-sight of the matter over-density, for a general redshift distribution of sources with $p=p(z)$ (see \citealt{bartelmann2001} and \citealt{schneider2005} for the underlying equations used in this section). This is given by
\begin{equation}
\label{convergence_def}
 \kappa \left(\bmath{\phi} \right)=\frac{3H_0^2\Omega_{\rm m}}{2c^2}\int_0^{\chi_H}d\chi \frac{\chi W\left( \chi\right)}{a\left( \chi\right)}\delta\left(\chi \bmath{\phi},\chi \right),
\end{equation}
where $W\left( \chi\right)$ is a weighting function (with $\chi$ being the co-moving distance) defined as 
\begin{equation}
\label{window_function_eq}
 W\left( \chi\right)=\int_{\chi}^{\chi_H}d\chi' G\left( \chi'\right)\frac{\chi' - \chi}{\chi'}
\end{equation}
with $G\left( \chi\right)$ given by $G\left( \chi\right)d\chi=p(z)dz$. In Equations (1) and (2), $c$ is the speed of
light, $\bmath{\phi}$ is the initial direction of the light
propagation, $\delta$ is the density contrast and $a=a\left(\chi \right)$
is the scale factor. Both the integrals extend up to the co-moving distance at the horizon, $\chi_H$. The function $p=p(z)$ describes how the SNe are distributed in redshift (see below for details).

The angular correlation of the convergence of two sources located at $\bmath{\phi}$ and $\bmath{\phi}+\bmath{\theta}$ (the modulus of the separation angle is $\theta$), is
\begin{equation}
\label{k_correlation}
 \langle \kappa \left(\bmath{\phi} \right)  \kappa \left(\bmath{\phi} + \bmath{\theta}\right) \rangle = \frac{1}{2\pi}\int_0^{\infty} lP_{\kappa}(l)J_0(l \theta)dl,
\end{equation}
where $P_{\kappa}(l)$ is the convergence power spectrum, as a function of the angular wave-number $l$, given by
\begin{equation}
\label{P_k}
 P_{\kappa}(l)=\frac{9H_0^4\Omega^2_{\rm m}}{4c^4}\int_0^{\chi_H} d\chi \frac{W^2\left(\chi \right)}{a^2\left(\chi \right)}P_{\delta}\left(\frac{l}{\chi},\chi \right).
\end{equation}
In Equation (\ref{P_k}), $P_{\delta}\left(\frac{l}{\chi},\chi \right)$ is
the total matter power spectrum, which is a function of both the Fourier mode
($k=l/\chi$) and time (via $\chi=\chi(t)$) and $J_0$ is the Bessel function of the first kind.
Equation (\ref{k_correlation}) is the correlation function we wish to detect.

The predicted noise on this correlation can be estimated from a simple expression of the Poisson noise. We do not include magnitude covariance in this first analysis, as off-diagonal terms are sub-dominant (see Section 3 for details). We also neglect the non-Gaussian fluctuations induced by lensing, as they are also sub-dominant \citep{marra2013}.
Under these assumptions, the expression for the noise becomes
\begin{equation}
\label{noise}
 \sigma_{\rm mm}=\frac{\sigma_{\rm err}^2}{\sqrt{N_{p}}},
\end{equation}
where $\sigma_{\rm mm}$ is the variance on the magnitude-magnitude correlation at a given angle, and $N_{\rm p}$ is the number of pairs with separation angle within $\left[\theta,\theta+\Delta \theta \right]$, given by
\begin{equation}
N_{\rm p}=\frac{N_{\rm s}(N_{\rm s}-1)}{2} \frac{2\pi \theta \Delta\theta}{A_{\rm s}},
\end{equation}
under the hypothesis that the $N_{\rm s}$ sources are uniformly distributed on the area $A_s$. 

In Equation (\ref{noise}), $\sigma_{\rm err}$ is the overall uncertainty on individual measurements of the SN magnitude. For this study, 
we allow $\sigma_{\rm err}$ to take values of 0.1, 0.15 and 0.2 magnitudes to cover the likely range of values for the intrinsic dispersion of SN magnitudes in current \citep{betoule2014} and future surveys. This replicates the approach taken in \cite{scovacricchi2016}. Within this range, the value $\sigma_{\rm err}=0.15$ is the most plausible number for the surveys considered in this paper \citep{bernstein2012,lsst_science_book}. We include $\sigma_{\rm err}=0.1$ mag in the analysis with the purpose of illustrating the potential of such measurements and the impact of a significant improvement to the SN standardization process, e.g. see \citealt{kelly2015} where they achieve $\sigma_{\rm err}\simeq0.07$ for a subset of SNeIa associated with young star-forming environments.

By combining Equations (\ref{k_correlation}) and (\ref{noise}), we can define the signal-to-noise ratio (SNR) as
\begin{equation}
  \label{SNR_eq}
 \text{SNR} = \left[ \frac{5}{\ln 10} \right]^2 \frac{\langle \kappa \left(\bmath{\phi} \right)  \kappa \left(\bmath{\phi} + \bmath{\theta}\right) \rangle}{\sigma_{\rm mm}},
\end{equation}
where the factor of $\left[5 / \ln 10 \right]^2$ converts the ratio from convergence into magnitude. 

The non-linear corrections to the matter power spectrum in Equation (\ref{P_k}) have been computed following the approach of \cite{smith2003} with the newly published values from \cite{takahashi2012}, starting from a linear power spectrum for adiabatic CDM with the transfer function by \cite{eisenstein1999}. We use the approximated growth factor from \cite{carroll1992}.

\subsection{Cross-correlation function}

In addition to studying the auto-correlation function, we can extend our study to include possible cross-correlations between supernova magnitudes and galaxies along the line-of-sight. The cross-power spectrum between the convergence and the density field is defined as
\begin{equation}
 P_{\rm \kappa \delta}(l) = \frac{3H_0^2\Omega_{\rm m}}{2c^2}\int_0^{\chi_H} d\chi \frac{W_{\rm SN}\left(\chi \right)G_{\rm gal}\left(\chi \right)}{\chi~a\left(\chi \right)}P_{\delta}\left(\frac{l}{\chi},\chi \right),
\end{equation}
where $W_{\rm SN}\left(\chi \right)$ comes from Equation (\ref{window_function_eq}), and contains the weighted redshift distribution of SNe, while $G_{\rm gal}\left(\chi \right)=p_{\rm gal}(z)(dz/d\chi)$ describes the galaxy redshift distribution. The function $p_{\rm gal}$ (normalized to unit area) is often parameterized using \citep{smail1995}
\begin{equation}
\label{p_gal}
 p_{\rm gal}(z) \propto z^{\alpha}\exp \left[ {-\left(\frac{z}{z_0}\right) ^\beta} \right],
\end{equation}
where $\alpha$, $\beta$ and $z_0$ are free parameters that can be adapted to describe different galaxy populations. This distribution has a peak at $z_{\rm peak}=\left( \frac{\alpha}{\beta}\right)^{1/\beta}z_0$, which reduces to $z_{\rm peak}\simeq z_0$ when $\alpha \simeq \beta$.
We can then compute the angular cross-correlation function by  
\begin{equation}
 \langle \kappa \left(\bmath{\phi} \right)  \delta \left(\bmath{\phi} + \bmath{\theta}\right) \rangle = \frac{1}{2\pi}\int_0^{\infty} lP_{\kappa \delta}(l)J_0(l \theta)dl,
\end{equation}
and the signal-to-noise ratio then becomes
\begin{equation}
 \rm SNR = \left[ \frac{5}{\ln 10} \right] \frac{b ~\langle \kappa \left(\bmath{\phi} \right)  \delta \left(\bmath{\phi} + \bmath{\theta}\right) \rangle}{\sigma_{\rm mg}}
\end{equation}
assuming a linear and deterministic bias $b$ between galaxies and the underlying distribution of matter (as was done, for instance, in \citealt{smith2014}), and $\sigma_{\rm mg}$ is the Poisson noise given by 
\begin{equation}
\sigma_{\rm mg} = \sigma_{\rm err} / \sqrt{N_{\rm cp}}
\end{equation}
(the cosmic variance is expected to be strongly subdominant). 
The number of SN-galaxy pairs within an annulus of mean radius $\theta$ is $N_{\rm cp}=N_{\rm s}\cdot (2\pi \Sigma_{\rm gal} \theta \Delta\theta)$, 
where $\Sigma_{\rm gal}$ is the surface density of galaxies on the sky. For simplicity, we fix the bias parameter $b=1$ throughout the paper.


\subsection{Survey description and parameters}

We aim to study the signal-to-noise predictions for the angular two-point correlation function for DES and LSST as the two major wide-field, distant SN surveys in the coming decade. In Figure \ref{DESandLSST_histograms}, we present the expected redshift histograms for both surveys, with the DES distribution obtained from \citet{bernstein2012}, split into the shallow and deep fields, and LSST normalized distribution from \citet{lsst_science_book}. The overall number of SNe assumed in both surveys (as a function of their wide and deep fields) is summarised in Table 1, e.g., for DES we assume 1850 SNe Ia over an area of 24 deg$^2$ (shallow survey) and 1650 SNe Ia over $6$ deg$^2$ (deep survey). 

For LSST, the expected number of SNe Ia remains uncertain, depending on the details of the final survey strategy in terms of the survey depth per epoch and in each passband. For the LSST wide survey (assuming 20000 deg$^2$), we assume three values for the total number of SNe Ia detected over the full 10 years of LSST operations. These values are given in Table 1 and are the same range of numbers presented in \citet{quartin2014}, so we can be consistent with this work. These assumed total numbers are consistent with the predictions given in Figure 11.6 (right-hand panels) of \cite{lsst_science_book}, which show that up to a million SNe Ia could be detected, depending on the signal-to-noise at maximum (SNRmax) in the light curve and the number of passbands assumed (although not all these SNe Ia will provide accurate distance estimates). For the LSST deep fields (assuming 7 fields, each of 10 deg$^2$), we again estimate a range for the total number of possible SNe Ia using Figure 11.6 of \cite{lsst_science_book} (left-hand panel). In this case, we use the predicted range seen in the number of SNe Ia per year for $N_{\rm fit}$(SNRmax)$=3$ with different SNRmax values (see Table 1).  

\begin{figure*}
\includegraphics[width=1\textwidth]{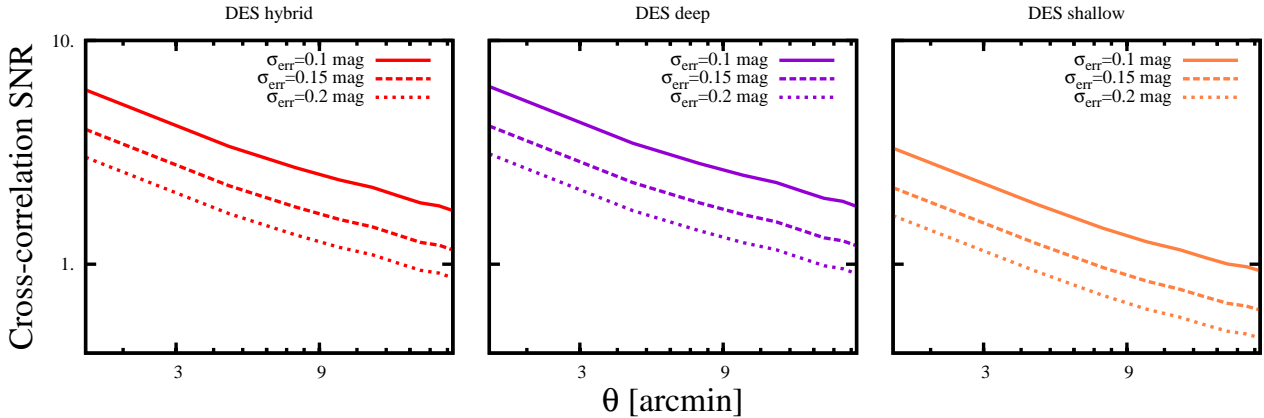}
\caption{The binned signal-to-noise predictions for DES cross-correlation function (bin size is 3 arcminutes) as a function of survey configuration, total number of SNe Ia and value of $\sigma_{\rm err}$ (see Table \ref{table_surveys} for details).}
\label{SNR_cross_DES}
\end{figure*}

\subsection{Analytical results}

For DES, we do not expect to detect the angular two-point auto-correlation of SN magnitudes for any reasonable combination of survey strategy and the three values of $\sigma_{\rm err}$ discussed above. At best, the DES SN auto-correlation function could reach SNR$\simeq2$ (on the smallest scales) with $\sigma_{\rm err}=0.1$ for the DES deep fields only. In Figure \ref{SNR_cross_DES}, we present our DES predictions for the SN-galaxy cross-correlation function, which has higher signal-to-noise values, due to the increased surface density of galaxies. For these predictions, we use Equation (9) for the DES galaxy redshift distribution, selecting values of $\alpha$, $\beta$ and $z_0$ that best represent the observed DES photometric redshift distribution (\citealt{giannantonio2015}; see Table \ref{table_galsurveys} for details). In Figure \ref{SNR_cross_DES}, we show the signal-to-noise as a function of separation, for both the deep and shallow fields combined and separately (the bias parameter is set to 1). The signal is dominated by the deep fields, which is expected given the increased redshift range probed and greater surface density of supernovae.

\begin{figure*}
\includegraphics[width=1\textwidth]{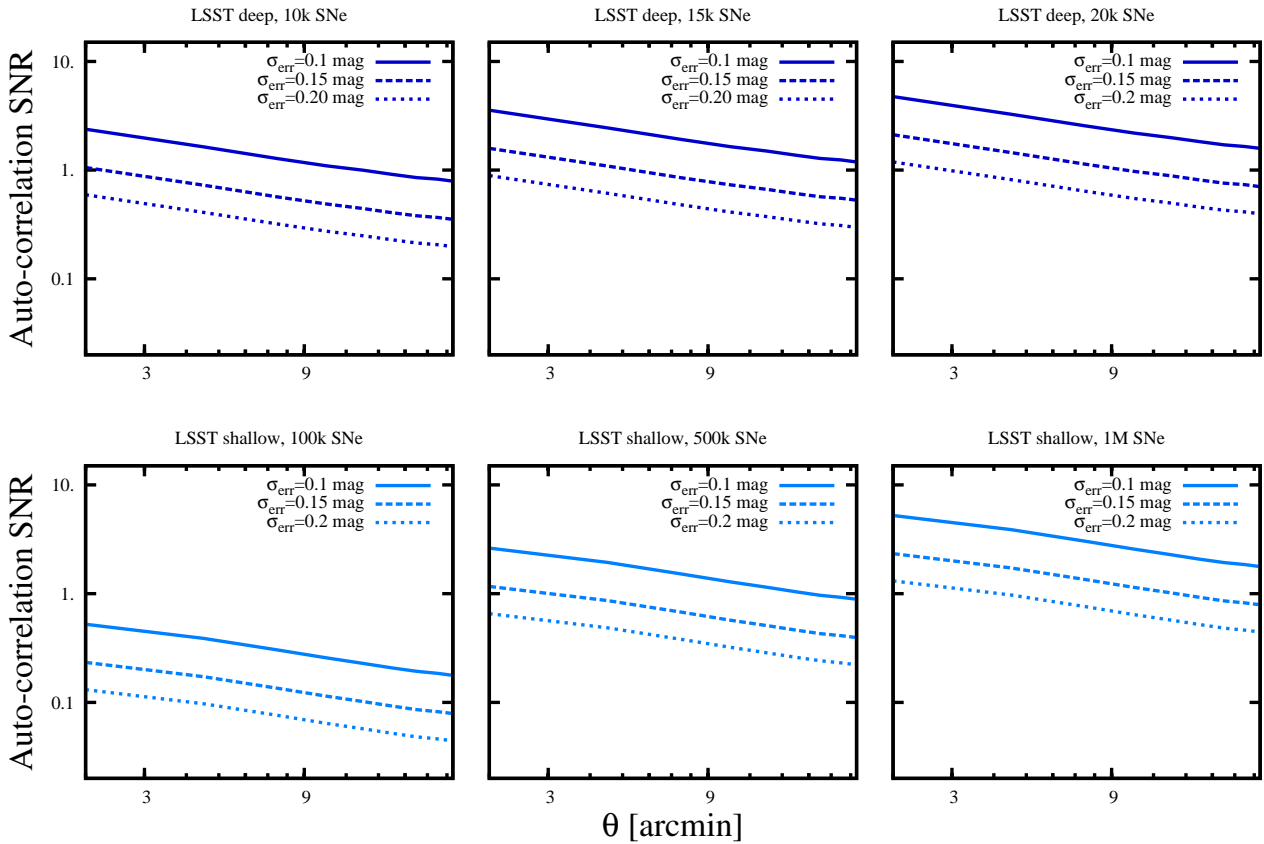}
\caption{The binned signal-to-noise predictions for LSST auto-correlation (bin size is 3 arcminutes) as a function of survey configuration, total number of SNe Ia and value of $\sigma_{\rm err}$ (see Table \ref{table_surveys} for details).}
\label{SNR_lsst}
\end{figure*}

In Figure \ref{SNR_lsst}, we show the predicted signal-to-noise for the auto-correlation as a function of angular separation for the LSST deep fields (top three panels) and LSST shallow survey (bottom panels). From left to right, we show the effect of increasing the total number of SNe Ia ($N_{s}$) as presented in Table \ref{table_surveys}. As expected, the overall signal-to-noise of the auto-correlation function increases with the number of SNe. This is most obvious for the LSST shallow survey populated with one million SNe, where even the least accurate SNe Ia distances (with $\sigma_{\rm err}=0.2$) provide a possible detection on the smallest angular scales. 
These predictions suggest that SN lensing will be detectable in the LSST wide survey, regardless of the overall quality of the light curves (and thus distance estimates), simply because of the high surface density of SNe available. For example, if we consider the $\sigma_{\rm err}=0.15$ case, which is the typical population scatter seen for present-day SN surveys, then we expect an integrated signal-to-noise of $\sim$5 on angular scales below 10 arcminutes (for the million LSST SNe Ia case). Such a measurement of the magnification lensing of SN would provide an excellent check of the LSST photometry and can be used to check systematic uncertainties associated with the galaxy shear weak lensing measurements \citep{semboloni2013}. The LSST deep fields should provide a robust detection of the SN lensing auto-correlation function. If we focus on the $\sigma_{\rm err}=0.1$ case (top right-hand panel), then we predict a total signal-to-noise of 7 integrated to 9 arcminutes (or a signal-to-noise of greater than two per bin for separations below 9 arcminutes). We can expect these LSST deep fields to deliver high signal-to-noise, well-sampled SNe Ia light curves, thus leading to more accurate SN distances more consistent with $\sigma_{\rm err}=0.1$. If such accurate distances cannot be achieved, then the SN lensing signal disappears quickly regardless of the total number of SNe Ia observed.

It is interesting to note the different dependences for our signal-to-noise predictions between the LSST shallow and deep surveys. For the shallow survey (bottom panels), the overall signal-to-noise does not change significantly with the total number of SNe considered, suggesting LSST wide will detect sufficient SNe Ia for such a lensing measurement. The opposite is true for the deep survey (top panels, Figure \ref{SNR_lsst}), where the quality of the SNe distances has the greatest effect. These measurements will therefore be complementary and could be combined to provide a significant SN lensing detection.

\begin{figure*}
\includegraphics[width=1\textwidth]{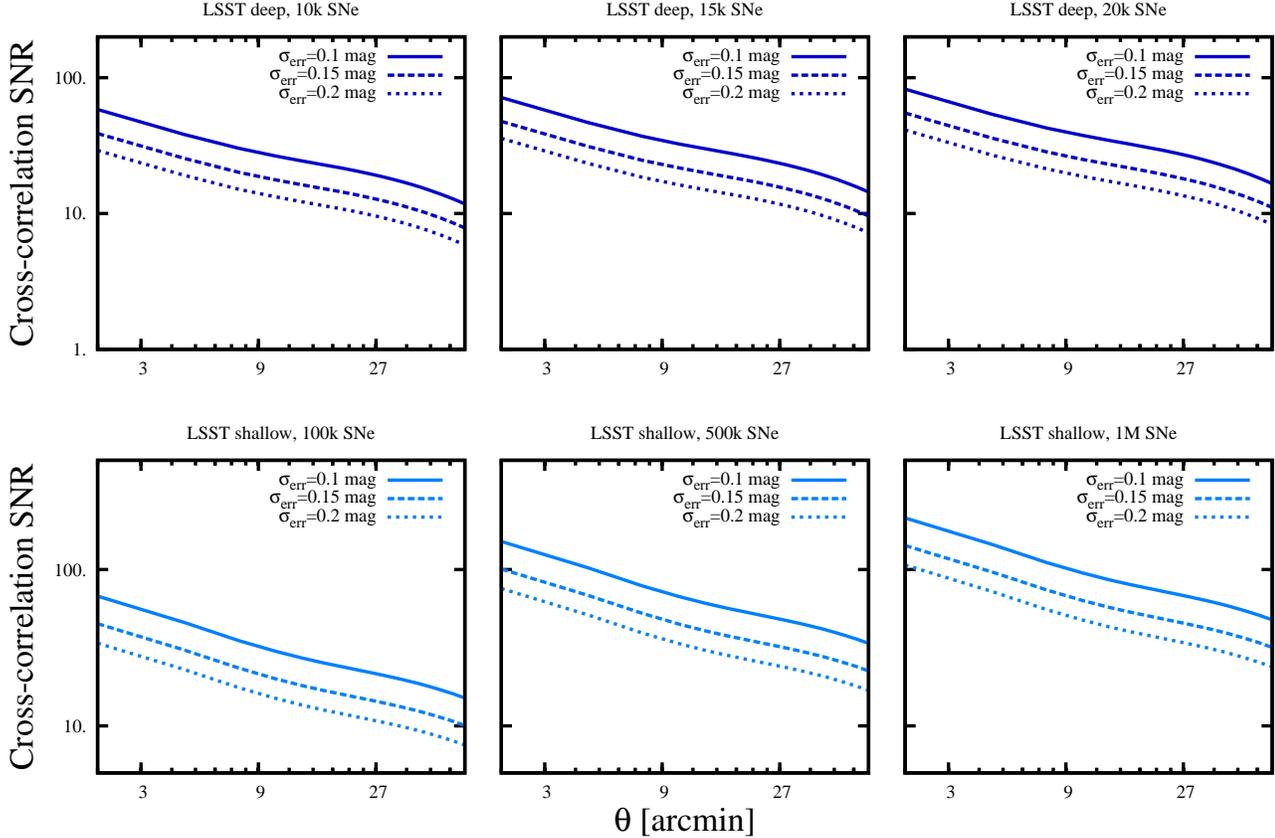}
\caption{The binned signal-to-noise predictions for LSST cross-correlation function (bin size is 3 arcminutes) as a function of survey configuration, total number of SNe Ia and value of $\sigma_{\rm err}$ (see Table \ref{table_surveys} for details).}
\label{SNR_cross_LSST}
\end{figure*}

In Figure \ref{SNR_cross_LSST}, we show the LSST SN-galaxy cross-correlation function for the same combination of survey configurations and SN numbers as presented in Figure \ref{SNR_lsst}. We provide in Table \ref{table_galsurveys} the assumed values of $\alpha$, $\beta$ and $z_0$ used in Equation (\ref{p_gal}) for the LSST galaxy redshift distribution \citep[see][Section 3.7.2 for details]{lsst_science_book}. This distribution reproduces the so-called LSST galaxy ``gold'' sample, which will include about four billion galaxies ($\Sigma_{\rm gal}=55$ gal/arcminutes$^2$) over 20,000 deg$^2$ of sky, observed with a high S/N (i $< 25.3$, corresponding to S/N$>$20 for point sources). The predicted signal is much more significant when correlating SNe with galaxies, regardless of our assumptions for $\sigma_{\rm err}$ and survey details. For example, a million SNe Ia with $\sigma_{\rm err}=0.2$ should deliver an integrated signal-to-noise of 1200 to 1 degree separations, providing an impressive cosmological probe for dark matter and dark energy studies. Considering the LSST deep fields, we expect an integrated SNR of 900 (up to 1 degree separation) for the intermediate case (15,000 SNe and $\sigma_{\rm err}=0.15$ mag).

\section{Simulations}
\label{section_mice}

Similar to \citet{macaulay2016}, we test our analytical predictions using mock SN catalogues created from the galaxy mock data of MICECAT v1.0\footnote{http://cosmohub.pic.es/}, which is the first public data release of the MICE Grand Challenge Simulation (MICE-GC). These mock catalogues are based on an N-body light-cone simulation, containing 70 billion dark matter particles in a $3 h^{-1} \text{Gpc}^3$ co-moving volume. From this dark matter simulation, halo and galaxy catalogues were created using a Halo Occupation Distribution and a Halo Abundance Matching technique as discussed in \citet{crocce2015}. In particular, we use a 100 deg$^2$ area extracted from the all-sky lensing map of \citet{fosalba2015}, built using the ``Onion Universe'' approach of \citet{fosalba2008}, which models the Universe as a set of concentric radial shells of finite width,  providing an estimate of convergence and shear on sub-arcminute scales. 
The fiducial cosmology for this simulation is a flat $\Lambda$CDM Universe with $\Omega_{\rm m}=0.25$, $\Omega_{\Lambda}=0.75$, $H_0=70$ Km s$^{-1}$ Mpc$^{-1}$ and $\sigma_8=0.8$. The MICE-GC Simulation does not include the effects of baryonic physics on the evolution of structures.


To create our mock SN catalogue, we randomly sub-sample from the galaxy MICECAT catalogue to obtain a set of SN host galaxies with the same redshift histograms as the DES and LSST deep field (as shown in Figure \ref{DESandLSST_histograms}). We divide the 100 deg$^2$ simulation area into three separate catalogues, namely 6 deg$^2$ with the redshift distribution of the DES deep survey, 24 deg$^2$ with the redshift distribution of the DES shallow survey and the remaining 70 deg$^2$ with the redshift distribution of the LSST deep survey (of the three values used in the previous section, we now select the intermediate value $N_{s}=15000$). We do not attempt to match the field configuration of these individual surveys; we simply match their total areal coverage and redshift distributions\footnote{This assumption may slightly over-estimate the SNR on small angular scales as a more realistic survey of several disjointed fields (like DES) would have fewer SN pairs on such scales. There would also be increased edge-effects. We expect these effects to be negligible compared to other uncertainties in our modelling.}.

Given a set of SNe for each survey, we must then create the Hubble residual for the $i^{\rm {th}}$ supernova of that dataset, which we define as
\begin{equation}
\label{hubble_res}
 \Delta {m}_{i} = \mu_{{\rm obs},i} - \mu_{\rm cos}({z}_{i})
\end{equation}
where $\mu_{\rm obs}$ and $\mu_{\rm cos}$ are the observed and best fit distance moduli respectively. For each supernova, the simulation provides the value of the convergence $\kappa_{i}\left(\bmath{\phi}_{i} \right)$ and the shear $\bmath{\gamma}_{i}\left(\bmath{\phi}_{i} \right)$ along that particular line of sight, $\bmath{\phi}_{i}$. Hence, we can directly compute the Hubble residual within the weak lensing approximation (including second order terms) from
\begin{equation}
\label{hubble_res_MICE}
 \Delta m_{i} = -2.5 \log \left[1+2\kappa_{i}+3\kappa_{i}^2+ |\bmath{\gamma}_{i}|^2 \right]+\delta m
\end{equation}
\citep{marra2013}. We include $\delta m$,  as an additional error term drawn at random from a Gaussian distribution with zero mean, but fixed width given by $\sigma_{\rm err}$. Cases with $\delta m=0$ are called ``lensing only'' and used for testing purposes.

We define the estimator for the SN magnitude two-point correlation function in the $k^{\rm {th}}$ angular bin as
\begin{equation}
\label{calc_corr}
 \langle \Delta m \Delta m \rangle (\bar{\theta_{k}}) = \sum_{\rm pairs} \frac{\Delta m_{i} \Delta m_{j}}{N_{\rm p}^{(k)}}
\end{equation}
where the sum extends over the $N_{\rm p}^{(k)}$ SN pairs with separation angle $\theta_{ij} \in [\theta_{k},\theta_{k}+\Delta \theta_{k} ]$. We also define $\bar{\theta_k}$ as the arithmetic average of the $\theta_{ij}$ in the $k^{\rm {th}}$ bin.
The angular separation of a pair of SNe is computed using the celestial coordinates (right ascension and declination) available from the simulations\footnote{The SN positions on the sky include lensing deflection effects. However, tests have shown us that these shifts do not affect our measurements.}.

To estimate errors, we randomly shuffled the values of $\Delta m_{i}$ across the SNe in our mock sample, keeping the angular positions of these SNe fixed, and the total number of SNe. We then repeated the correlation function measurement using Equation (\ref{calc_corr}). From this, we obtain a 
series of measurements $\xi_k^{(r)}=\langle \Delta m \Delta m \rangle^{(r)} (\bar{\theta_k})$ where the index $k$ represents the angular bin and the index $r$ represents the different shuffled data-sets ($r=1,...,N_{\rm r}$ and $N_{\rm r}$ the number of trials). We then define the covariance matrix of the data as
\begin{equation}
\label{covariance_matrix}
 C_{ij} = \langle (\xi_{i}^{(r)}  -  \bar{\xi}_{i} )(\xi_{j}^{(r)} - \bar{\xi}_{j} )\rangle
\end{equation}
averaged over the $N_{\rm r}$ measurements. 

For each measurement (given by Equation \ref{calc_corr}), we associate an error bar of $\sigma_k=(C_{kk})^{\frac{1}{2}}$, explicitly
\begin{equation}
\label{corr_errbars}
 \sigma_k= \sqrt{\frac{\sum_{r=1}^{N_{\rm r}} \left( \langle \Delta m \Delta m \rangle^{(r)} (\bar{\theta_k}) - \overline{\langle \Delta m \Delta m \rangle} (\bar{\theta_k}) \right)^2}{N_{\rm r}}}
\end{equation}
where the average $ \overline{\langle \Delta m \Delta m \rangle}$ is
\begin{equation}
 \overline{\langle \Delta m \Delta m \rangle}(\bar{\theta_k})=\sum_{r=1}^{N_{\rm r}} \frac{\langle \Delta m \Delta m \rangle^{(r)} (\bar{\theta_k})}{N_{\rm r}}.
\end{equation}
We verify that $N_{\rm r}=500$ provides sufficient trials to achieve a stable estimate of the diagonal terms of the covariance matrix (we neglect the non-diagonal elements, since they are at least one order of magnitude smaller). 

We also fit our mock correlation functions to the cosmological predictions introduced in Section \ref{analitycal_predictions}. In order to do this, we use the $\chi^{2}$ statistic to simultaneously fit for $\Omega_{\rm m}$ and $\sigma_8$ using 
\begin{equation}
\label{chi2_eq}
 \chi^2 = \left(\bmath{\xi}_{\rm d}-\bmath{\xi}_{\rm t}\right)^{\rm t} C ^{-1} \left(\bmath{\xi}_{\rm d}-\bmath{\xi}_{\rm t} \right)
\end{equation}
where $\bmath{\xi}_{\rm d}$ and $\bmath{\xi}_{\rm t}$ are respectively the array of data and theoretical values (computed from Equation \ref{k_correlation} for any $\theta_{i}$ of interest) and $C$ is the (diagonal) covariance matrix from Eq. (\ref{covariance_matrix}). 

\begin{figure}
\includegraphics[width=\columnwidth]{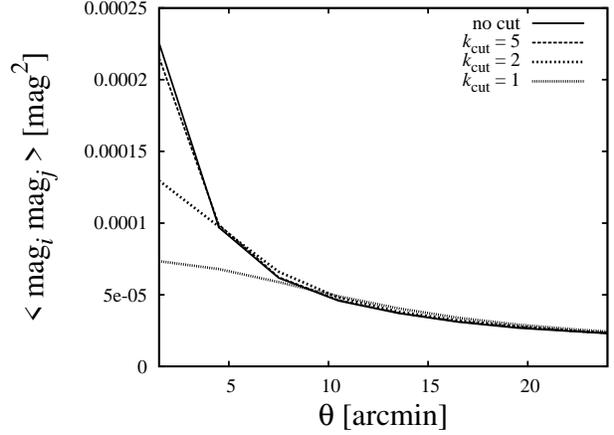}
\caption{Magnitude correlation function (Eq. $\ref{k_correlation}$) when different cuts in power are applied ($P_\delta(k)\equiv0$ for $k>k_{\rm cut}$). The specific values of $k_{\rm cut}$ are reported in the legend.}
\label{plot_PS}
\end{figure}

A consideration when comparing our correlation function measurements to theoretical predictions is the contribution of the small scale power to the auto-correlation function at a given angle. To assess this, we present in Figure \ref{plot_PS} our theoretical SN correlation function as a function of separation angle, for different values of $k_{\rm cut}$, which is the cut-off scale applied to the matter power spectrum in the integral of Equation \ref{k_correlation} (i.e., $P_\delta (k)\equiv0~$ for $k>k_{\rm cut}$). Figure \ref{plot_PS} shows that the correlation function is sensitive to scales up to 5 $h$/Mpc, while the contribution of smaller scales (higher values of $k$) is negligible. However in this work we do not apply any cut-off to our integrations.

\begin{figure}
\includegraphics[width=\columnwidth]{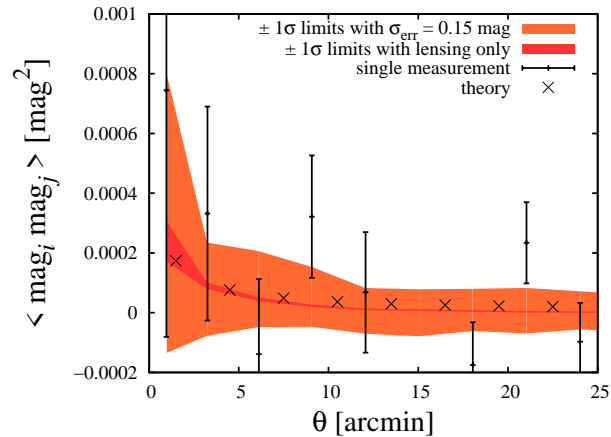}
\caption{Auto-correlation results for DES hybrid strategy, for a single mock realisation (data points) and $\pm 1 \sigma$ contours for 100 realisations of the same field (the red region is obtained selecting $\sigma_{\rm err}=0$, the orange region with $\sigma_{\rm err}=0.15$ mag). Crosses are computed via Equation (\ref{k_correlation}), within the same fiducial cosmology specified in Section 3.}
\label{plot_measurements_DES}
\end{figure}

\begin{figure*}
\includegraphics[width=.45\textwidth]{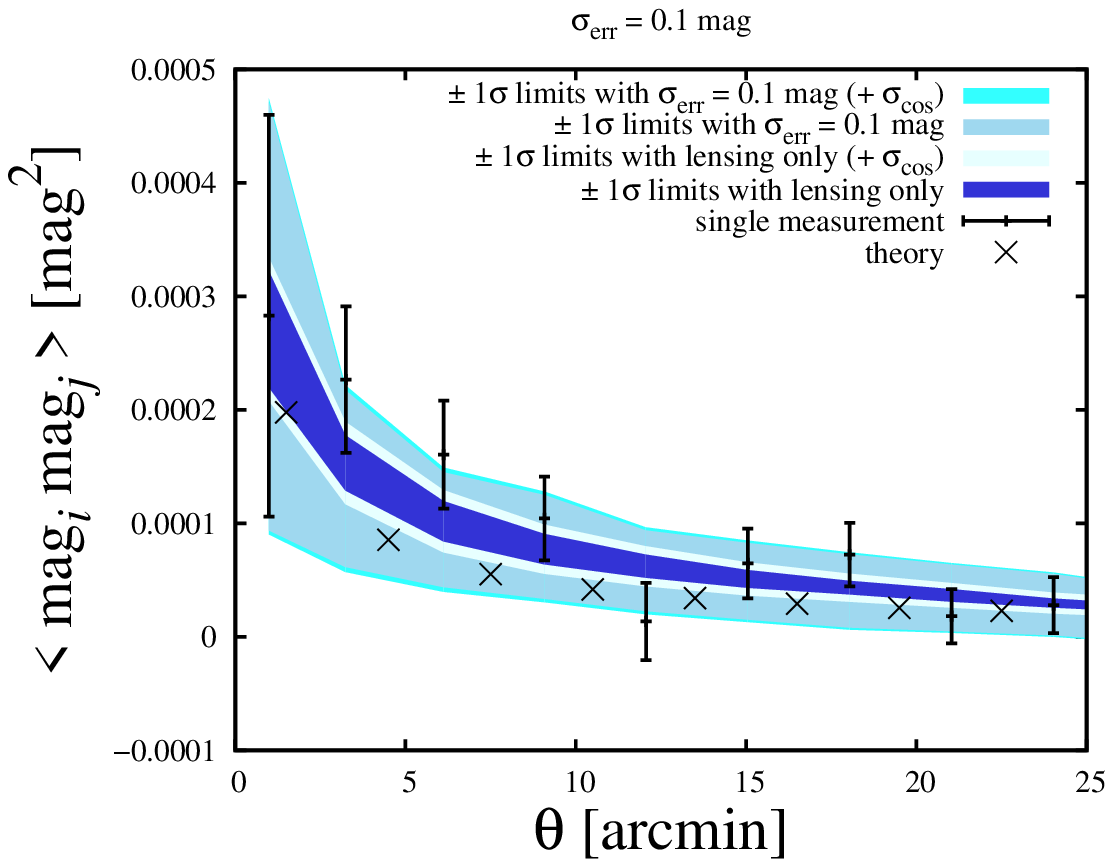}
\includegraphics[width=.45\textwidth]{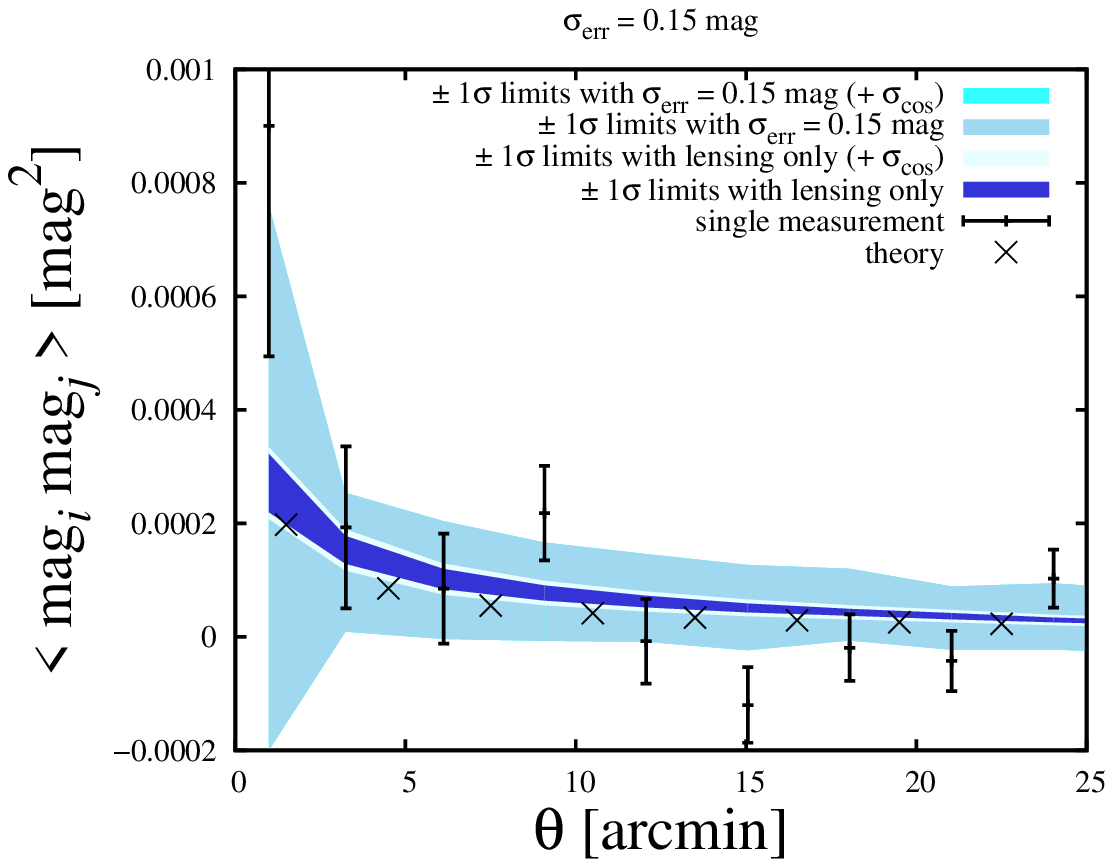}
\caption{Auto-correlation results from simulations. Results for LSST deep, for a single mock realisation (data points) and $\pm 1 \sigma$ contours for 100 realisations of the same field. The blue regions are obtained selecting $\sigma_{\rm err}=0$, while sky blue regions with $\sigma_{\rm err}=0.10$ mag or $\sigma_{\rm err}=0.15$ mag  respectively for the left and right panel. Square points are computed via Equation (\ref{k_correlation}), within the same fiducial cosmology specified in Section 3. The contour widening due to cosmic variance is highlighted by the light-cyan (lensing only) and cyan contours.}
\label{plot_measurements_LSST}
\end{figure*}

In Figures \ref{plot_measurements_DES} and \ref{plot_measurements_LSST} we show auto-correlation results from our mock SN catalogues for both the DES and LSST deep survey. In order to reduce the impact of the sampling noise when presenting results and allow a direct comparison with analytical results of Section \ref{analitycal_predictions}, we show the measurements from a single realisation
of the fields (as described earlier in this section) alongside the confidence intervals for 100 realisations of the same survey (the confidence intervals have been computed as plus-or-minus the standard deviation on 100 statistically equivalent realisations of the same Hubble diagram). 

On Figure \ref{plot_measurements_DES} we show the 1$\sigma$ confidence region of 100 realisations of the DES hybrid HD, when selecting $\sigma_{\rm err}=0$ (i.e. lensing only - red area) and $\sigma_{\rm err}=0.15$ mag (orange area). We also show on the same plot the data-points of the measurements for a single mock realisation (as described in 3.1), when $\sigma_{\rm err}=0.15$ mag. This result shows that we will be unable to detect the SN magnitude auto-correlation with DES, confirming what was previously found via the analytical method.

On Figure \ref{plot_measurements_LSST} we show the results for the LSST deep survey, using 15k SNe distributed on an area of 70 deg$^2$. Also in this case, we show the contours for 100 realisations of the same field using $\sigma_{\rm err}=0$ (lensing only in the label, blue area) and $\sigma_{\rm err}=0.10$ mag or 0.15 mag (sky blue area) respectively for the panel on the left and on the right. 
For these specific cases, we also include on the contour widths (adding it in quadrature) an estimate of the cosmic variance, which represents a possible source of systematic uncertainty when measuring physical quantities within a small area of the sky (as in this case, 70 deg$^2$). Details of the procedure are given in the Appendix, while the effects on the 1$\sigma$ contours (from multiple realisations of the LSST HD) are shown in Figure \ref{plot_measurements_LSST} (``+ $\sigma_{\rm cos}$'' in the legend). As usual, we distinguish between the lensing only case (light-cyan contours) and the one with $\sigma_{\rm err}\neq 0$ (cyan contours). As expected, the cosmic variance significantly influences the lensing only contours (being $\sigma_{\rm cos}\sim 10^{-5}$ mag$^2$), while it is completely negligible when introducing the intrinsic scatter on the SN distance moduli. The blue contours of Figure \ref{plot_measurements_LSST} (both panels) also highlight a disagreement between the theoretical prediction for the angular correlation and the lensing only multiple realisation contours (blue area). To check whether or not this is a systematic effect, we repeat the procedure for 12 statistically equivalent LSST patches, for which we take the sample variance (per bin) as an estimation of the cosmic variance (detail in Appendix). Being the disagreement outlined in Figure \ref{plot_measurements_LSST} of the same order as $\sigma_{\rm cos}$ ($\sim 10^{-5}$ mag$^2$), we then conclude that this difference is within the uncertainties caused by cosmic variance.

The data points shown in Figure \ref{plot_measurements_LSST} are the measurements from single realisations (cosmic variance is not included, as negligible) of the LSST HD. The case with $\sigma_{\rm err}=0.1$ mag (l.h.s.) leads us to an SNR$\simeq 18$ for the first 10 angular bins, consistent with the analytical result shown in Figure \ref{SNR_lsst} (top-central panel).  Both panels of Figure \ref{plot_measurements_LSST} also show the theoretical correlation, computed via Equation (\ref{k_correlation}) and within the same fiducial cosmology as the MICE simulation.
By applying Eq. (\ref{chi2_eq}) to the same set of data points, we find a $\chi^2$ per degree of freedom $\chi_{\rm dof}^2 \simeq 1.6$ for the fiducial cosmology assumed in the MICE simulation (for the first 10 angular bins), while assuming a null correlation (i.e. $\bmath{\xi}_{\rm t} \equiv 0, ~ \forall ~ \theta_k$) gives $\chi^2_{\rm dof} \simeq 5.0$. We also compute the signal-to-noise ratio for a single mock realisation of the LSST deep survey, now selecting $\sigma_{\rm err}=0.15$ mag (shown on the right panel of Figure \ref{plot_measurements_LSST}), finding SNR$\simeq$ 6 (integrated on the first 10 angular bins.) and $\chi_{\rm dof}^2 \simeq 1.7$, when comparing the data-points with the fiducial values. This decrease is expected, due to the high sensitivity of such measurement to the intrinsic dispersion on the SN magnitudes and compatible with the analytical results shown in the previous section.

\section{Discussion and Conclusions}
\label{conclusions}

In this paper, we have considered the possibility of detecting angular correlations in the magnitudes for present and future supernova surveys. As shown in Section 2 and confirmed in Section 3, we do not expect the auto-correlation of these SN magnitudes to be significant in current surveys like DES. This conclusion is based on both analytical predictions and numerical simulations (which agree) and supported by an attempt to measure such angular SN auto-correlations with the existing JLA sample (see Appendix B). These findings are consistent with previous attempts to find evidence for SN lensing in the SDSS \citep{jonsson2008,smith2014} and SNLS \citep{kronborg2010,jonsson2010} samples.

In Figures \ref{SNR_cross_DES} and \ref{SNR_cross_LSST}, we show how it will be possible to detect SN magnitude correlations using the SN-galaxy angular cross-correlation function. In this way, we can significantly decrease the shot noise from the finite number of SNe in our survey, potentially leading to a $\simeq 15\sigma$ detection of SN lensing in the DES deep fields alone (integrated up to 24 arcminutes, $\sigma_{\rm err}=0.15$). Such a measurement would be an excellent test of the DES photometric calibration and provide additional cosmological constraints beyond the standard SN Hubble diagram fit. 

We also studied the measurement of the SN correlation functions assuming possible LSST SN surveys. In these cases, both the auto and cross-correlation functions should deliver clear detections of the SN lensing magnification signal, providing an additional probe of the dark matter distribution beyond studies of the moments of the SN magnitude residuals \citep{marra2013,quartin2014,castro2014,Amendola2015,castro2015,macaulay2016}. 

\begin{figure*}
\includegraphics[width=.48\textwidth]{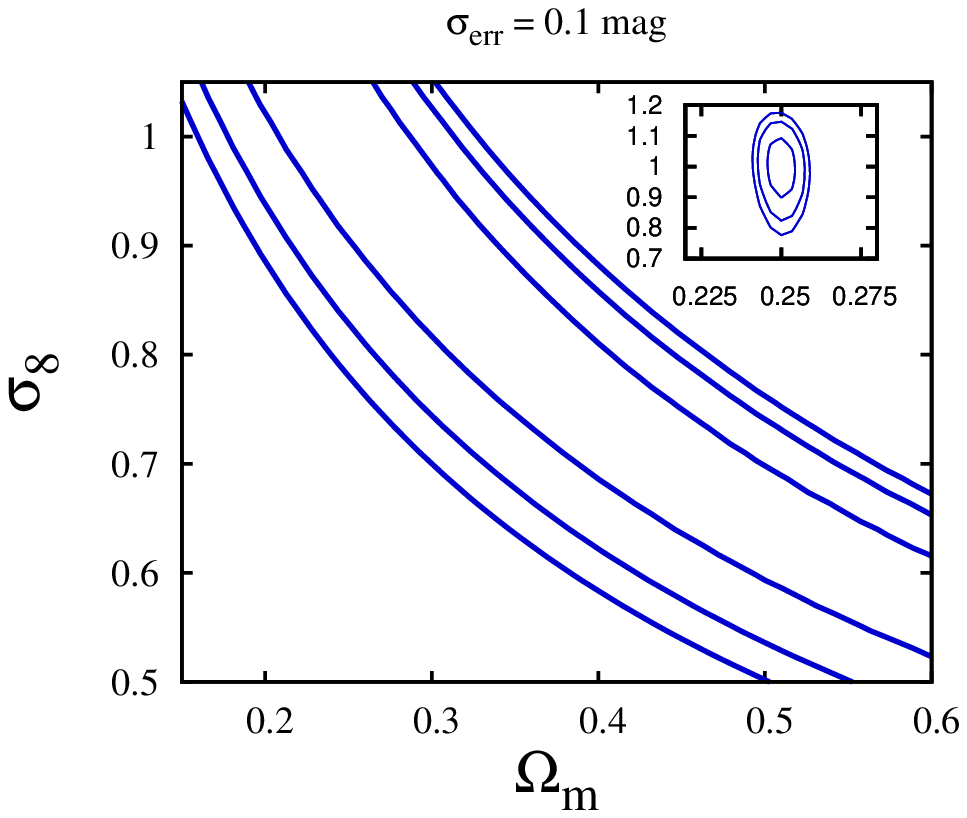}
\includegraphics[width=.48\textwidth]{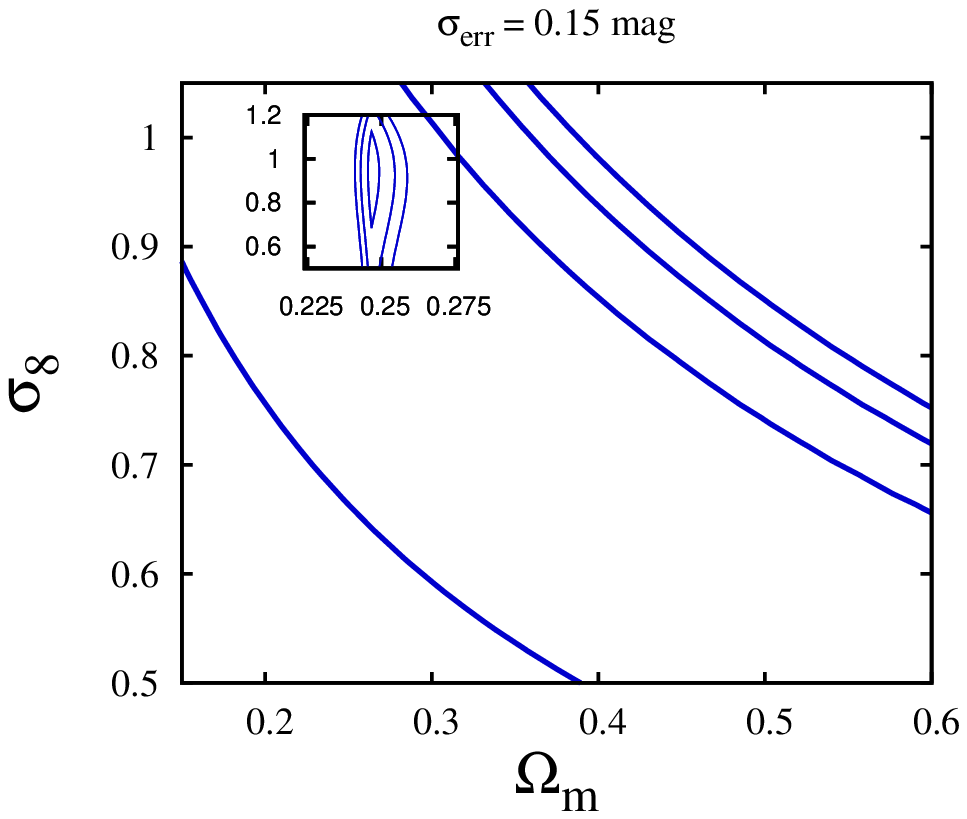}
\caption{68\%, 95\% and 99\% confidence level contours on the $\Omega_{\rm m}$--$\sigma_8$ plane when fitting the auto-correlation function of a single mock of the LSST deep survey (15k SNe on 70 deg$^2$) with $\sigma_{\rm err}=0.1$ mag (left) and $\sigma_{\rm err}=0.15$ mag (right). \emph{Inset} Same contours when a Gaussian prior on $\Omega_{\rm m}$ is applied, of width $\sigma_{\Omega_{\rm m}}=0.003$ (0.004)  when a value $\sigma_{\rm err}=0.1$ ($\sigma_{\rm err}=0.15$) is selected.}
\label{plot_fitcos}
\end{figure*}

To illustrate the potential of auto-correlation measurements, we present in Figure \ref{plot_fitcos} the 1, 2 and 3$\sigma$ contours\footnote{The contours have been computed selecting the contour levels of the $\chi^2$ from its minimum, as described in \cite{press1993}.} (respectively at 68\%, 95\% and 99\% confidence levels) for $\Omega_{\rm m}$ and $\sigma_8$ using the auto-correlation of the mock SN data obtained from the single realisation of LSST deep ($\sigma_{\rm err}=0.1$/0.15 mag, left/right panel) using Equation \ref{chi2_eq} (the data points are shown in Figure \ref{plot_measurements_LSST}). We limit our analysis to the first ten angular bins (i.e. $\theta \leqslant 30 $ arcmin).

This result with $\sigma_{\rm err}=0.1$ mag (l.h.s.) is comparable with the constraints found by \citealt{quartin2014} (their Figure 5 pag. 8, red contours) when fitting the lensing distribution moments (second to fourth) using a simulated dataset for LSST with 100,000 SNe Ia and $\sigma_{\rm err}=0.12$ mag. We stress that we did not include information about geometry, usually recovered by fitting the Hubble diagram and which is able to constrain $\Omega_{\rm m}$ and $w$, the Dark Energy equation of state parameter. We can easily assume that, by the time we have LSST 10 year data, we certainly have very tight constraints on the matter density parameter. Hence, the parameter fitting of the angular correlation would reduce to $\sigma_8$ only. 

To include this additional piece of information, we also show on the top right corner of Figure \ref{plot_measurements_LSST} (l.h.s.) the contour levels when applying a Gaussian prior on $\Omega_{\rm m}$, with standard deviation $\sigma_{\Omega_{\rm m}}=0.003$. The width of the prior has been computed by fitting the mock HD for LSST deep with a custom-made likelihood code\footnote{This simple likelihood code fits for $\Omega_{\rm m}$ only after marginalizing for the unknown amplitude of the Hubble diagram with analytical procedures. All the details can be found in \citep{scovacricchi2016}.} as explained in \cite{scovacricchi2016}. Projecting the contours onto the $\sigma_8$ axis, we find $\sigma_{\sigma_8} \simeq 0.1$ which is competitive with the contours shown in \citet{quartin2014}, who found $\sigma_{\sigma_8} \simeq 0.06$ by fitting the moments of the magnification distribution function (first to fourth) on a mock LSST data sample composed by 100k SNe Ia and $\sigma_{\rm err}=0.12$ mag (Figure 5 pag. 8, central panel). 
Repeating this procedure for the same LSST deep mock HD with increased intrinsic scatter ($\sigma_{\rm err}=0.15 $ mag, shown on the right hand panel of Figure \ref{plot_fitcos}) reveals again the sensitivity of the magnitude correlation function to the value of $\sigma_{\rm err}$. The $\Omega_{\rm m}$ - $\sigma_8$ contours are broader (see Figure \ref{plot_fitcos}, right panel) and the broadening magnitude is expected if we consider that the data-points of Figure \ref{plot_measurements_LSST} clearly show an upper-limit on the strength of the auto-correlation (which will reflect on strong constraints in the area of the parameter space with high $\Omega_{\rm m}$ and $\sigma_8$), while the lower-limit is weak. The resulting likelihood surface for the two cosmological parameters is then very asymmetric. This is further confirmed by the contours computed by applying the gaussian prior on the matter density parameter (now $\sigma_{\Omega_{\rm m}}=0.004$), where we find $\sigma_{\sigma_8}=0.2$ at 1$\sigma$ level of confidence.

In conclusion, in this paper we have shown two possible applications of the Type Ia SN Hubble diagram beyond the usual cosmological parameter fit. We have applied analytical methods to predict future weak lensing applications within DES and LSST surveys, finding that a cross correlation between foreground galaxies and SN Hubble residuals will deliver clear detections for both surveys. We have also found that the study of the auto-correlation of SN magnitudes is more difficult than the cross-correlation counter-part, due to the lower SN surface density with respect to the galaxy surface density. For this reason, using both analytical methods and numerical simulations, we confirm that we do not expect that a future measurement of the mag-mag auto-correlation will be possible with the Dark Energy Survey, unless the SN intrinsic scatter will be reduced to values smaller than 0.1 mag. 

The same methods lead us to the conclusion that a similar measurement will be of interest for the LSST, for which we have simulated future HDs and found that we can both detect this signal and use it to constrain $\Omega_{\rm m}$ and $\sigma_8$ with an accuracy comparable to the one predicted using the weak lensing one-point statistics within the same survey, if the current standardization processes will be improved to achieve $\sigma_{\rm err}=0.1$ mag. We also found the auto-correlation measure to be very sensitive to the value of the intrinsic scatter. This suggests that for the future of the non-standard HD analyses presented here, the improvement of this value will be vital.

Both the detection of the weak lensing features on the HD and its cosmological fit will be important for non-standard applications of SN cosmology (possibly joined with weak lensing one-point statistics), not only providing new independent measurements of the cosmological parameters, but also important checks for SN calibration and their related systematics in the upcoming SN wide-area surveys (e.g. differences in the photometric calibration of supernovae located in separated fields could be highlighted by a SN magnitude cross-correlation between those patches). The "one-point" and the "two-point" weak lensing correlations are complementary methodologies and their joined analysis will be an interesting step beyond the usual Hubble diagram fit.

\section*{Acknowledgements}
We would like to thank Tamara Davis for useful comments on a draft of this paper.  DS thanks the Faculty of Technology of the University of Portsmouth for support during his PhD studies. DB, EM and RCN acknowledge funding from STFC grant ST/N000668/1 and EM acknowledges the Australian Research Council Centre of Excellence for All-sky Astrophysics (CAASTRO), through project number CE110001020. 


\bibliographystyle{mn2e}

\bibliography{Bibliography}

\begin{thebibliography}{}

\bibitem[\protect\citeauthoryear{Amendola, Castro, Marra \& Quartin}{Amendola
  et~al.}{2015}]{Amendola2015}
Amendola L.,  Castro T.,  Marra V.,    Quartin M.,  2015, Mon. Not. Roy.
  Astron. Soc., 449, 2845

\bibitem[\protect\citeauthoryear{{Bartelmann} \& {Schneider}}{{Bartelmann} \&
  {Schneider}}{2001}]{bartelmann2001}
{Bartelmann} M.,  {Schneider} P.,  2001, \physrep, 340, 291

\bibitem[\protect\citeauthoryear{{Bernstein}, {Kessler}, {Kuhlmann}, {Biswas},
  {Kovacs}, {Aldering}, {Crane}, {D'Andrea}, {Finley} \& {Frieman}}{{Bernstein}
  et~al.}{2012}]{bernstein2012}
{Bernstein} J.~P.,  {Kessler} R.,  {Kuhlmann} S.,  {Biswas} R.,  {Kovacs} E.,
  {Aldering} G.,  {Crane} I.,  {D'Andrea} C.~B.,  {Finley} D.~A.,    {Frieman}
  2012, \apj, 753, 152

\bibitem[\protect\citeauthoryear{{Betoule}, {Kessler}, {Guy}, {Mosher},
  {Hardin}, {Biswas}, {Astier} \& {El-Hage}}{{Betoule}
  et~al.}{2014}]{betoule2014}
{Betoule} M.,  {Kessler} R.,  {Guy} J.,  {Mosher} J.,  {Hardin} D.,  {Biswas}
  R.,  {Astier} P.,    {El-Hage} 2014, \aap, 568, A22

\bibitem[\protect\citeauthoryear{{Carroll}, {Press} \& {Turner}}{{Carroll}
  et~al.}{1992}]{carroll1992}
{Carroll} S.~M.,  {Press} W.~H.,    {Turner} E.~L.,  1992, \araa, 30, 499

\bibitem[\protect\citeauthoryear{{Castro}, {Marra} \& {Quartin}}{{Castro}
  et~al.}{2016}]{castro2016}
{Castro} T.,  {Marra} V.,    {Quartin} M.,  2016, \mnras

\bibitem[\protect\citeauthoryear{{Castro} \& {Quartin}}{{Castro} \&
  {Quartin}}{2014}]{castro2014}
{Castro} T.,  {Quartin} M.,  2014, \mnras, 443, L6

\bibitem[\protect\citeauthoryear{Castro, Quartin \& Benitez-Herrera}{Castro
  et~al.}{2015}]{castro2015}
Castro T.,  Quartin M.,    Benitez-Herrera S.,  2015

\bibitem[\protect\citeauthoryear{{Cooray}, {Holz} \& {Huterer}}{{Cooray}
  et~al.}{2006}]{cooray2006_signal}
{Cooray} A.,  {Holz} D.~E.,    {Huterer} D.,  2006, \apjl, 637, L77

\bibitem[\protect\citeauthoryear{{Cooray}, {Huterer} \& {Holz}}{{Cooray}
  et~al.}{2006}]{cooray2006_noise}
{Cooray} A.,  {Huterer} D.,    {Holz} D.~E.,  2006, Physical Review Letters,
  96, 021301

\bibitem[\protect\citeauthoryear{{Crocce}, {Castander}, {Gazta{\~n}aga},
  {Fosalba} \& {Carretero}}{{Crocce} et~al.}{2015}]{crocce2015}
{Crocce} M.,  {Castander} F.~J.,  {Gazta{\~n}aga} E.,  {Fosalba} P.,
  {Carretero} J.,  2015, \mnras, 453, 1513

\bibitem[\protect\citeauthoryear{{Dodelson} \& {Vallinotto}}{{Dodelson} \&
  {Vallinotto}}{2006}]{dodelson2005}
{Dodelson} S.,  {Vallinotto} A.,  2006, \prd, 74, 063515

\bibitem[\protect\citeauthoryear{{Eisenstein} \& {Hu}}{{Eisenstein} \&
  {Hu}}{1999}]{eisenstein1999}
{Eisenstein} D.~J.,  {Hu} W.,  1999, \apj, 511, 5

\bibitem[\protect\citeauthoryear{{Fosalba}, {Gazta{\~n}aga}, {Castander} \&
  {Crocce}}{{Fosalba} et~al.}{2015}]{fosalba2015}
{Fosalba} P.,  {Gazta{\~n}aga} E.,  {Castander} F.~J.,    {Crocce} M.,  2015,
  \mnras, 447, 1319

\bibitem[\protect\citeauthoryear{{Fosalba}, {Gazta{\~n}aga}, {Castander} \&
  {Manera}}{{Fosalba} et~al.}{2008}]{fosalba2008}
{Fosalba} P.,  {Gazta{\~n}aga} E.,  {Castander} F.~J.,    {Manera} M.,  2008,
  \mnras, 391, 435

\bibitem[\protect\citeauthoryear{{Giannantonio}, {Fosalba}, {Cawthon}, {Omori},
  {Crocce}, {Elsner}, {Leistedt}, {Dodelson}, {Benoit-L{\'e}vy},
  {Gazta{\~n}aga}, {Holder}, {Peiris} \& {Percival}}{{Giannantonio}
  et~al.}{2016}]{giannantonio2015}
{Giannantonio} T.,  {Fosalba} P.,  {Cawthon} R.,  {Omori} Y.,  {Crocce} M.,
  {Elsner} F.,  {Leistedt} B.,  {Dodelson} S.,  {Benoit-L{\'e}vy} A.,
  {Gazta{\~n}aga} E.,  {Holder} G.,  {Peiris} H.~V.,    {Percival} W.~J.,
  2016, \mnras, 456, 3213

\bibitem[\protect\citeauthoryear{{Hui} \& {Greene}}{{Hui} \&
  {Greene}}{2006}]{hui2006}
{Hui} L.,  {Greene} P.~B.,  2006, \prd, 73, 123526

\bibitem[\protect\citeauthoryear{{J{\"o}nsson}, {Kronborg}, {M{\"o}rtsell} \&
  {Sollerman}}{{J{\"o}nsson} et~al.}{2008}]{jonsson2008}
{J{\"o}nsson} J.,  {Kronborg} T.,  {M{\"o}rtsell} E.,    {Sollerman} J.,  2008,
  \aap, 487, 467

\bibitem[\protect\citeauthoryear{{J{\"o}nsson}, {Sullivan}, {Hook}, {Basa},
  {Carlberg}, {Conley} \& {Fouchez}}{{J{\"o}nsson} et~al.}{2010}]{jonsson2010}
{J{\"o}nsson} J.,  {Sullivan} M.,  {Hook} I.,  {Basa} S.,  {Carlberg} R.,
  {Conley} A.,    {Fouchez} 2010, \mnras, 405, 535

\bibitem[\protect\citeauthoryear{{Kelly}, {Filippenko}, {Burke}, {Hicken},
  {Ganeshalingam} \& {Zheng}}{{Kelly} et~al.}{2015}]{kelly2015}
{Kelly} P.~L.,  {Filippenko} A.~V.,  {Burke} D.~L.,  {Hicken} M.,
  {Ganeshalingam} M.,    {Zheng} W.,  2015, Science, 347, 1459

\bibitem[\protect\citeauthoryear{{Kessler}, {Marriner}, {Childress},
  {Covarrubias}, {D'Andrea}, {Finley}, {Fischer}, {Foley}, {Goldstein}, {Gupta}
  \& {Kuehn}}{{Kessler} et~al.}{2015}]{kessler2015}
{Kessler} R.,  {Marriner} J.,  {Childress} M.,  {Covarrubias} R.,  {D'Andrea}
  C.~B.,  {Finley} D.~A.,  {Fischer} J.,  {Foley} R.~J.,  {Goldstein} D.,
  {Gupta} R.~R.,    {Kuehn} 2015, \aj, 150, 172

\bibitem[\protect\citeauthoryear{{Kronborg} \& {Hardin}}{{Kronborg} \&
  {Hardin}}{2010}]{kronborg2010}
{Kronborg} T.,  {Hardin} 2010, \aap, 514, A44

\bibitem[\protect\citeauthoryear{{Linder}, {Wagoner} \& {Schneider}}{{Linder}
  et~al.}{1988}]{linder1988}
{Linder} E.~V.,  {Wagoner} R.~V.,    {Schneider} P.,  1988, \apj, 324, 786

\bibitem[\protect\citeauthoryear{{LSST Science Collaboration}, {Abell},
  {Allison}, {Anderson}, {Andrew}, {Angel}, {Armus}, {Arnett}, {Asztalos},
  {Axelrod} \& et al.}{{LSST Science Collaboration}
  et~al.}{2009}]{lsst_science_book}
{LSST Science Collaboration} {Abell} P.~A.,  {Allison} J.,  {Anderson} S.~F.,
  {Andrew} J.~R.,  {Angel} J.~R.~P.,  {Armus} L.,  {Arnett} D.,  {Asztalos}
  S.~J.,  {Axelrod} T.~S.,    et al. 2009, ArXiv e-prints

\bibitem[\protect\citeauthoryear{{Macaulay}, {Davis}, {Scovacricchi}, {Bacon},
  {Collett} \& {Nichol}}{{Macaulay} et~al.}{2016}]{macaulay2016}
{Macaulay} E.,  {Davis} T.~M.,  {Scovacricchi} D.,  {Bacon} D.,  {Collett}
  T.~E.,    {Nichol} R.~C.,  2016, ArXiv e-prints

\bibitem[\protect\citeauthoryear{{Marra}, {Quartin} \& {Amendola}}{{Marra}
  et~al.}{2013}]{marra2013}
{Marra} V.,  {Quartin} M.,    {Amendola} L.,  2013, \prd, 88, 063004

\bibitem[\protect\citeauthoryear{{Perlmutter}, {Aldering}, {Goldhaber}, {Knop},
  {Nugent}, {Castro}, {Deustua}, {Fabbro}, {Goobar}, {Groom} \&
  {Hook}}{{Perlmutter} et~al.}{1999}]{perl99}
{Perlmutter} S.,  {Aldering} G.,  {Goldhaber} G.,  {Knop} R.~A.,  {Nugent} P.,
  {Castro} P.~G.,  {Deustua} S.,  {Fabbro} S.,  {Goobar} A.,  {Groom} D.~E.,
  {Hook} 1999, \apj, 517, 565

\bibitem[\protect\citeauthoryear{{Planck Collaboration}, {Ade}, {Aghanim},
  {Armitage-Caplan}, {Arnaud}, {Ashdown}, {Atrio-Barandela}, {Aumont},
  {Baccigalupi}, {Banday} \& et al.}{{Planck Collaboration}
  et~al.}{2014}]{planck_collaboration}
{Planck Collaboration} {Ade} P.~A.~R.,  {Aghanim} N.,  {Armitage-Caplan} C.,
  {Arnaud} M.,  {Ashdown} M.,  {Atrio-Barandela} F.,  {Aumont} J.,
  {Baccigalupi} C.,  {Banday} A.~J.,    et al. 2014, \aap, 571, A16

\bibitem[\protect\citeauthoryear{{Quartin}, {Marra} \& {Amendola}}{{Quartin}
  et~al.}{2014}]{quartin2014}
{Quartin} M.,  {Marra} V.,    {Amendola} L.,  2014, \prd, 89, 023009

\bibitem[\protect\citeauthoryear{{Riess}, {Filippenko}, {Challis},
  {Clocchiatti}, {Diercks} \& {Garnavich}}{{Riess} et~al.}{1998}]{riess98}
{Riess} A.~G.,  {Filippenko} A.~V.,  {Challis} P.,  {Clocchiatti} A.,
  {Diercks} A.,    {Garnavich} 1998, \aj, 116, 1009

\bibitem[\protect\citeauthoryear{{Sako}, {Bassett}, {Becker}, {Brown},
  {Campbell}, {Cane}, {Cinabro}, {D'Andrea}, {Dawson}, {DeJongh}, {Depoy} \&
  {Dilday}}{{Sako} et~al.}{2014}]{sako2014}
{Sako} M.,  {Bassett} B.,  {Becker} A.~C.,  {Brown} P.~J.,  {Campbell} H.,
  {Cane} R.,  {Cinabro} D.,  {D'Andrea} C.~B.,  {Dawson} K.~S.,  {DeJongh} F.,
  {Depoy} D.~L.,    {Dilday} 2014, ArXiv e-prints

\bibitem[\protect\citeauthoryear{Schneider, Meylan, Kochanek, Jetzer, North \&
  Wambsganss}{Schneider et~al.}{2006}]{schneider2005}
Schneider P.,  Meylan G.,  Kochanek C.,  Jetzer P.,  North P.,    Wambsganss
  J.,  2006, Gravitational Lensing: Strong, Weak and Micro: Saas-Fee Advanced
  Course 33.
Saas-Fee Advanced Course, Springer Berlin Heidelberg

\bibitem[\protect\citeauthoryear{{Scovacricchi}, {Nichol}, {Bacon}, {Sullivan}
  \& {Prajs}}{{Scovacricchi} et~al.}{2016}]{scovacricchi2016}
{Scovacricchi} D.,  {Nichol} R.~C.,  {Bacon} D.,  {Sullivan} M.,    {Prajs} S.,
   2016, \mnras, 456, 1700

\bibitem[\protect\citeauthoryear{{Semboloni}, {Hoekstra}, {Huang}, {Cardone},
  {Cropper}, {Joachimi}, {Kitching}, {Kuijken}, {Lombardi}, {Maoli}, {Mellier},
  {Miller}, {Rhodes}, {Scaramella}, {Schrabback} \& {Velander}}{{Semboloni}
  et~al.}{2013}]{semboloni2013}
{Semboloni} E.,  {Hoekstra} H.,  {Huang} Z.,  {Cardone} V.~F.,  {Cropper} M.,
  {Joachimi} B.,  {Kitching} T.,  {Kuijken} K.,  {Lombardi} M.,  {Maoli} R.,
  {Mellier} Y.,  {Miller} L.,  {Rhodes} J.,  {Scaramella} R.,  {Schrabback} T.,
     {Velander} M.,  2013, \mnras, 432, 2385

\bibitem[\protect\citeauthoryear{{Smail}, {Ellis}, {Fitchett} \&
  {Edge}}{{Smail} et~al.}{1995}]{smail1995}
{Smail} I.,  {Ellis} R.~S.,  {Fitchett} M.~J.,    {Edge} A.~C.,  1995, \mnras,
  273, 277

\bibitem[\protect\citeauthoryear{{Smith}, {Bacon}, {Nichol}, {Campbell},
  {Clarkson}, {Maartens}, {D'Andrea}, {Bassett}, {Cinabro}, {Finley},
  {Frieman}, {Galbany}, {Garnavich}, {Olmstead}, {Schneider}, {Shapiro} \&
  {Sollerman}}{{Smith} et~al.}{2014}]{smith2014}
{Smith} M.,  {Bacon} D.~J.,  {Nichol} R.~C.,  {Campbell} H.,  {Clarkson} C.,
  {Maartens} R.,  {D'Andrea} C.~B.,  {Bassett} B.~A.,  {Cinabro} D.,  {Finley}
  D.~A.,  {Frieman} J.~A.,  {Galbany} L.,  {Garnavich} P.~M.,  {Olmstead}
  M.~D.,  {Schneider} D.~P.,  {Shapiro} C.,    {Sollerman} J.,  2014, \apj,
  780, 24

\bibitem[\protect\citeauthoryear{{Smith}, {Peacock}, {Jenkins}, {White},
  {Frenk}, {Pearce}, {Thomas}, {Efstathiou} \& {Couchman}}{{Smith}
  et~al.}{2003}]{smith2003}
{Smith} R.~E.,  {Peacock} J.~A.,  {Jenkins} A.,  {White} S.~D.~M.,  {Frenk}
  C.~S.,  {Pearce} F.~R.,  {Thomas} P.~A.,  {Efstathiou} G.,    {Couchman}
  H.~M.~P.,  2003, \mnras, 341, 1311

\bibitem[\protect\citeauthoryear{{Takahashi}, {Sato}, {Nishimichi}, {Taruya} \&
  {Oguri}}{{Takahashi} et~al.}{2012}]{takahashi2012}
{Takahashi} R.,  {Sato} M.,  {Nishimichi} T.,  {Taruya} A.,    {Oguri} M.,
  2012, \apj, 761, 152

\bibitem[\protect\citeauthoryear{{Teukolsky}, {Vetterling}, {Flannery}, {Lloyd}
  \& {Rees}}{{Teukolsky} et~al.}{1993}]{press1993}
{Teukolsky} S.~A.,  {Vetterling} W.~T.,  {Flannery} B.~P.,  {Lloyd} C.,
  {Rees} P.,  1993, The Observatory, 113, 214

\end{thebibliography}

\begin{figure}
\includegraphics[width=\columnwidth]{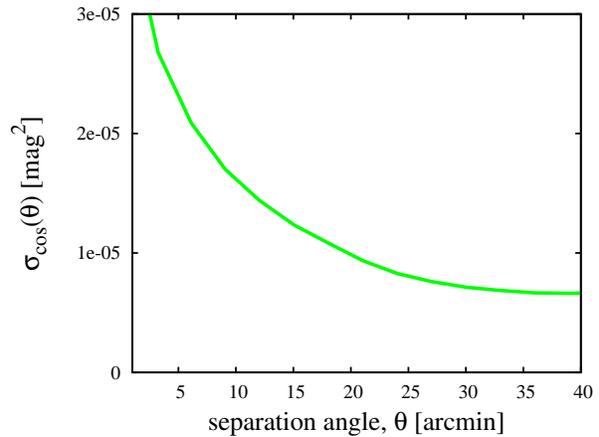}
\caption{Square root of the cosmic variance ($\sigma_{\rm cos}$) affecting the magnitude auto-correlation measurements, as a function of separation angle. This quantity is computed on 12 contiguous LSST-deep patches.}
\label{plot_cosmic_variance}
\end{figure}

\appendix
\section{Assessing the cosmic variance}
To assess the impact of cosmic variance when measuring the weak lensing auto-correlation function, we make use of the MICECAT simulated catalogue again. From a simulated area of $28 \times 30 $ deg$^2$ we cut 12 contiguous patches of sky, each one of $70$ deg$^2$. Within each patch, we randomly sample 100 HDs following the LSST deep survey redshift distribution (as given in Figure \ref{DESandLSST_histograms}) 
and considering 15,000 SNe. We then apply Eq. (\ref{calc_corr}) to each Hubble diagram, measuring the auto-correlation over 3 arcminute wide angular bins. 
We take the cosmic variance ($\sigma_{\rm cos}^2$) to be the square of the standard deviation (per angular bin) of the 12 patches' centroid values of the auto-correlation function (computed as the average on the 100 HD realisations), when no intrinsic scatter is selected.  We find $\sigma_{\rm cos}(\theta)$ to be a decreasing function of the angle and of the order of $\sim 10^{-5}$ mag$^2$ for angles below 40 arcminutes (see Figure \ref{plot_cosmic_variance}). Hence $\sigma_{\rm cos}$ is comparable on arcminute scale angles with the width of the lensing only contours (blue, Figure \ref{plot_measurements_LSST}, $\sigma_{\rm err}= 0.1$ mag), while is negligible when compared with the contours that include the intrinsic scatter. The  square root of the cosmic variance is also negligible in comparison to the errorbars given by Eq. \ref{corr_errbars} (shown in Figure \ref{plot_measurements_LSST}, ``single measurement'').

\section{JLA Results}

\begin{figure}
\centering
\includegraphics[width=\columnwidth]{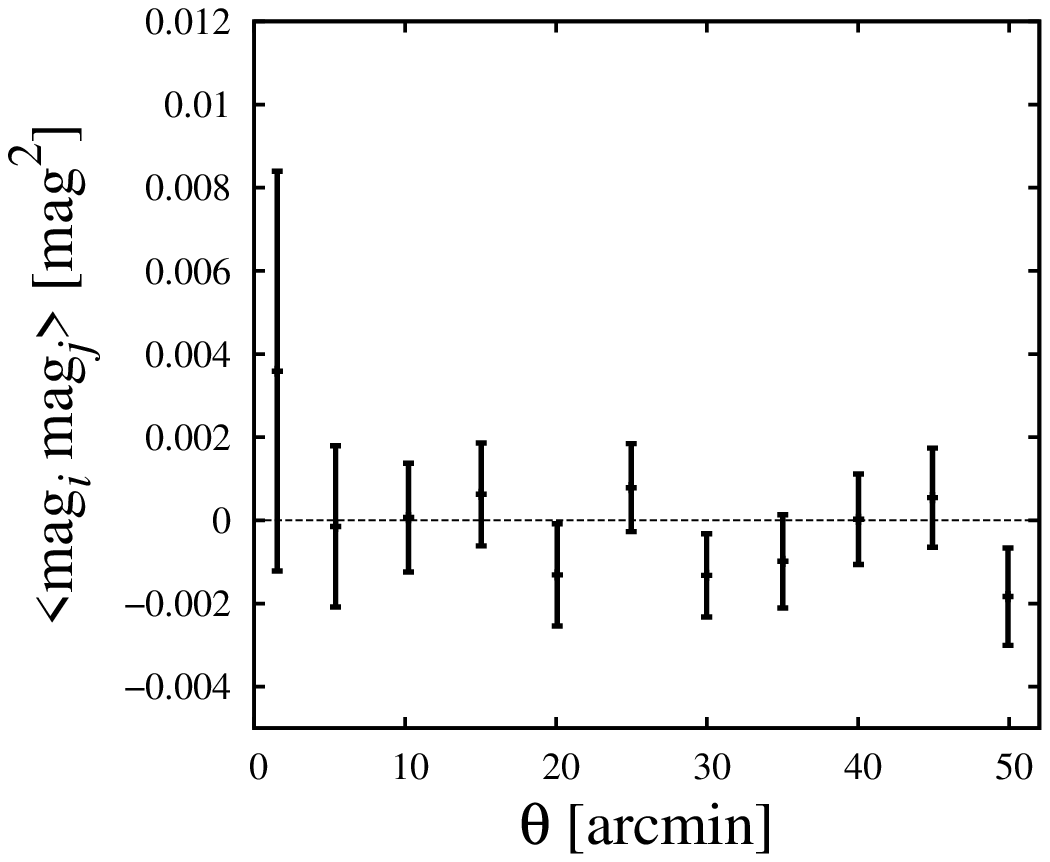}
\caption{JLA results for the mag-mag correlation function, using the whole set of 740 SNe. Bin width is 5 arcminutes.}
\label{plot_jla_only}
\end{figure}

Here we apply the method described in Section \ref{section_mice} to the 740 SNe Ia of the Jointed Light-curve Analysis (JLA) dataset \citep{betoule2014}\footnote{The dataset is publicly available at http://supernovae.in2p3.fr .}, composed of the full three years of SDSS, SNLS, HST as well as several nearby surveys. Re-adapting the notation of \cite{betoule2014}, the observed distance modulus for the i-th SN of the sample can be written as
\begin{equation}
\label{jla_distmod_obs}
 \mu_{{\rm obs},i} = m_{{\rm B},i} - \left(M_{\rm B}^{1} + \Delta_{\rm M} - \alpha \cdotp X_{1,i} + \beta \cdotp C_{i} \right)
\end{equation}
where $m_{\rm B}$, $X_{1}$ and $C_{1}$ are the observed peak magnitude in rest-frame B band, the time-stretching parameter for the light-curve and the color of the $i^{\rm {th}}$ supernova.
$M_{\rm B}^{1}$, $\Delta_{\rm M}$, $\alpha$, $\beta$ are the nuisance parameters of the fit, respectively the absolute magnitude of the SN, the step parameter that accounts for the observed correlation between the SN magnitudes and the mass of the host galaxies, and two nuisance parameters for the stretch and color corrections. 
The cosmological distance modulus is (for a comoving distance in Mpc)
\begin{equation}
\label{jla_distmod_th}
 \mu_{\rm cos}= 5 \log \left[ (1+z_{\rm hel})\chi(z_{\rm cmb},\Omega_{\rm m})\right]+25
\end{equation}
and it is a function of the matter density parameter $\Omega_{\rm m}$ (once the Hubble parameter $H_0$ is fixed to the value 70 Km/s/Mpc in a flat FRW universe) and the heliocentric and CMB rest frame redshifts. Using the published values of the best-fit parameters\footnote{The best fit parameter values are collected in \cite{betoule2014}, table 10 page 16 (row JLA, stat+sys)}, we can compute the Hubble residual for each SN by combining Equations (\ref{jla_distmod_obs}) and (\ref{jla_distmod_th}) via Eq. (\ref{hubble_res}). We compute the angular correlation function in the k-th bin using a weighted average, to account for the different uncertainties
on the SN magnitude measurements,
\begin{equation}
\label{w_average}
  \langle \Delta m \Delta m \rangle (\bar{\theta_{k}}) = \frac{\sum_{\rm pairs} w_{ij} \Delta m_{i} \Delta m_{j}}{\sum_{\rm pairs} w_{ij}}
\end{equation}
where the weights $w_{ij}=\left(\sigma_{m_i}^2+\sigma_{m_j}^2 \right)^{-1/2}$ and $\sigma_{m_{i,j}}$ are the peak magnitude uncertainties. Equation (\ref{w_average}) returns Eq. (\ref{calc_corr}) in the limit of equal weights. We then apply Equation (\ref{corr_errbars}) to estimate the errorbars of the correlation measurements. As expected, due to the small number of objects composing the JLA data sample as well as their distribution across the sky, we do not detect any correlation (see Figure \ref{plot_jla_only}).

\end{document}